\documentstyle[psfig,12pt]{article}

\def\beq{\begin{equation}}
\def\eeq{\end{equation}}

\def\be{\begin{equation}}
\def\bea{\begin{eqnarray}}
\def\ee{\end{equation}}
\def\eea{\end{eqnarray}}

\def\eqref#1{(\ref{#1})}

\def\a{\alpha}

\def\bra{\langle}
\def\ket{\rangle}
\def\e{{\rm e}}
\def\tr{{\rm tr}}

\begin{document}


\begin{titlepage}

\begin{centering}

\vspace*{3cm}

{\Large\bf ${\cal N}=(1,1)$ Super Yang--Mills on a (2+1)
Dimensional Transverse Lattice with one Exact Supersymmetry}

\vspace*{1.5cm}

{\bf  Motomichi Harada and Stephen Pinsky}
\vspace*{0.5cm}

{\sl Department of Physics \\
Ohio State University\\
Columbus OH 43210}

\vspace*{0.5cm}

\vspace*{1cm}


\vspace*{1cm}

\vspace*{1cm}

\begin{abstract}
We present a formulation of ${\cal N}=(1,1), $ Super Yang--Mills theory in 2+1 dimensions using
a transverse lattice methods that exactly preserves one supersymmetry. First, using a Lagrangian
approach we obtain a standard transverse lattice formulation of the Hamiltonian. We then show
that the Hamiltonian also can be written discretely as the square of a supercharge and
that this produces a different result. 
Problems associated with the discrete realization of the full supercharge
algebra are discussed. Numerically we solve  for the bound states of the theory in
the large
$N_c$ approximation  and  we find good convergence. We show that the massive
fermion and boson bound states are all exactly degenerate and that the number
of fermion and boson massless bound states are closely related. Also we
find that this theory admits winding states in the transverse direction and
that their masses vary inversely with the winding number. 
\end{abstract}
\end{centering}

\vfill

\end{titlepage}
\newpage

\section{Introduction}
In 1976 Bardeen and Pearson \cite{bardeen,BPR80} proposed formulating a quantum field
theory with a subset of the dimensions  discretized on a spatial lattice. In the
discrete spatial directions the theory was constructed to have discrete gauge
invariance, identical to conventional lattice gauge
theory. The remaining dimensions were then left to be treated by some other
method.  It is only in the last few years however that this idea has been fully
exploited. There are two directions of research which  rely on this idea that
have been of some interest in recent years. 

One research direction 
simply goes by the name ``transverse lattice" \cite{dv}. For a review see
\cite{rev}. This is a numerical method for solving  quantum field theory.
The name ``transverse lattice" is somewhat deceptive because the method is
actually the combination of several independent ideas, of which transverse
lattice is only one. The other ingredient has to do with how one treats the longitudinal
directions. There are analytical approaches to the longitudinal part of the
problem designed to carefully treat the end points in momentum space \cite{rev},
however they greatly limit the size of the basis one might use. The more common method
goes by the name discrete light-cone quantization (DLCQ), and was itself proposed
in 1985 by Brodsky and Pauli
\cite{pb85} as  a numerical method to solve quantum field
theories\cite{bpp98}.  In DLCQ one quantizes the theory with a Hamiltonian
$P^-=(P^0-P^1)/\sqrt{2}$ that evolves the system in light-cone time
$x^+=(x^0+x^1)/\sqrt{2}$, uses the light-cone gauge $A^+=0$, and places the system
in a light-cone spatial box. Thus, in this version of the transverse
lattice approach one treats the longitudinal directions with a discrete momentum
lattice and the transverse directions with a discrete spatial lattice. 

The other research direction that uses the transverse lattice focuses on
theories of extra dimensions. The directions that are put on a discrete
lattice are dimensions that are beyond the normal 3+1 dimensions of
conventional field theory. This approach suggested by Arkani-Hamed \cite
{ham} and Hill
\cite{Hill:2000mu} goes by the name ``deconstruction". Again the name is not
very descriptive. The point is that when the extra dimensions are put on a
discrete spatial lattice the extra dimensional field theory takes the form of
a (3+1) dimensional theory where each of the fields carries additional labels
corresponding to the structure of the extra dimensional space. If one is
creative about this extra dimensional space this method can be a mechanism
for constructing new and interesting field theories. Furthermore, since the
theory is formulated as a field theory in 3+1 dimensions the renormalization
is controllable.

With this very brief background we would like to suggest a slightly different
direction. We have recently been studying a supersymmetric formulation of DLCQ
which we call SDLCQ \cite{Antonuccio:1998kz,Antonuccio:1998jg,alpt98}. In many ways this
approach is similar to DLCQ, however, it is formulated in such a way that the theory, which
has discrete momenta and cutoffs in momentum space, is exactly supersymmetric.
For a review,  see \cite{Lunin:1999ib}.
Exact supersymmetry brings a number of very important numerical advantages to the method.
In 1+1 and 2+1 \cite{Antonuccio:1999zu,Haney:2000tk,hpt2001} dimensions
supersymmetric field theories are finite. We have also seen greatly improved
convergence. In this paper we will attempt to formulate a (2+1)--dimensional
${\cal N}=1$ Super Yang--Mills (SYM) theory as a SDLCQ theory in 1+1 dimensions
with a transverse spatial lattice in the one transverse direction. The challenge
is to formulate it such that it is supersymmetric exactly at every order of the
numerical approximation.

We will not be able to fully realize this goal. There are
several fundamental problems that prevent complete success. In formulating this
theory with gauge invariance in the one transverse dimension the gauge field
is replaced by a complex unitary link field.  Within the context of DLCQ
this field is quantized as a linear complex field. This then disturbs the
supersymmetry which usually requires the same number of fundamental fermion and
boson fields. In some sense this is a restatement of the error we are making by
treating a unitary field as a general complex field. There are simply too
many boson degrees of freedom relative to the number of fermion degrees of
freedom.  Conventionally one adds a potential to a transverse lattice theory to
enforce the unitarity of this complex boson field, but this  is not
possible within the context of an exactly supersymmetric theory. 
However, in the formulation of Gauss's law on the transverse lattice, one
finds that color conservation must be enforced at every lattice site. This
greatly reduces the number of allowed boson degrees of freedom. It is unclear
however if this constraint is sufficient to reduce the number of boson
degrees of freedom to the number required by unitarity.

We will be able to partially formulate SDLCQ for this theory and write the
Hamiltonian as the square of a supercharge. Previously we considered this
situation in a different class of theories \cite{Lunin:2000im}.  We will show
that this produces a different and simpler discrete Hamiltonian than the
standard lagrangian formulation.   When we solve this theory using this partial
formulation of SDLCQ we find that all of the massive states have exact
fermion-boson degeneracy as required by full supersymmetry. Our partial SDLCQ
does not require degeneracy between the massless fermion and boson states. We
find however that they are nearly equal in number. The solution can be viewed
as a unitary transformation from the constrained basis to an unconstrained
basis and we see that in this new basis the number of fermion and boson degrees
of freedom are very nearly equal. In effect, the partial supersymmetry and
Gauss's law are sufficient to approximately enforce the same symmetry in the spectrum that we
would have obtained had we been able to enforce unitarity. Recently Dalley and
Van de Sande \cite{Dalley:2001gj,Dalley:2002nj} have also pointed out the
importance of Lorentz symmetry in enforcing the constraint of unitarity. 

Since color is conserved at every transverse lattice site, there are two fundamentally different
types of states. For one class of states the color flux winds around the space one
or more times. We refer to these as cyclic states and to the other class of states as
non-cyclic states. The spectrum for both classes of states are presented. For the cyclic states 
we present the spectrum as a function of the number of windings.

In Section 2, we present the standard lagrangian formulation of this
theory of adjoint fermions and adjoint bosons. We show that Hamiltonian is
sixth order in the field. In Section 3, we present the SDLCQ formulation which
turns out to be only fourth order in the field. We show that there are two types
of allowed states. One type loops the entire transverse space, and we study
these state in Section 4. The states of the other type are localized, and we study
these states in Section 5.  In section 6 we discuss our conclusions and future
work.
\section{Transverse Lattice Model in 2+1 Dimensions} 
\label{sec:translat}
In this section we present the standard formulation of a transverse lattice
model in 2+1 dimensions of an ${\cal N}=(1,1)$ supersymmetric $SU(N_c)$ theory with
{\it both} adjoint bosons and adjoint fermions in the large--$N_c$ limit.

We work in light cone coordinates so that $x^{\pm}\equiv (x^0\pm
x^1)/\sqrt 2$. The metric is specified by $x^\pm=x_\mp$ and $x^2=-x_2$. Suppose that there
are
$N_{sites}$ sites in the transverse direction $x^2$ with lattice
spacing $a$. With each site, 
$i$,  we associate one gauge boson field $A_{\nu,i}(x^{\mu})$ and one spinor
field
$\Psi_i(x^{\mu})$, where $\nu, \mu=\pm$. $A_{\nu,i}$'s and $\Psi_i$'s are in the
adjoint representation. The adjacent sites, say $i$ and $i+1$, are connected by
what we call the link variables
$M_i(x^{\mu})$ and $M_i^{\dag}(x^{\mu})$, where $M_i(x^{\mu})$ stands for a link which
goes from the $i$-th site to the $(i+1)$-th site and $M_i^{\dag}(x^{\mu})$ for a link
from the $(i+1)$-th to the $i$-th site. We impose the periodic condition on the
transverse sites so that $A_{N_{sites}+1}=A_1$, $\Psi_{N_{sites}+1}=\Psi_1$,
$M_{N_{sites}+1}=M_1$ and $M^{\dag}_{N_{sites}+1}=M^{\dag}_1$. Under the transverse
gauge transformation \cite{rev} the fields transform as 
\begin{equation}
gA_i^{\mu} \longrightarrow U_igA_i^{\mu}U_i^{\dag}-iU_i\partial^{\mu}
  U_i^{\dag}, \quad
  M_i\longrightarrow U_iM_iU_{i+1}^{\dag}, \quad
  \Psi_i \longrightarrow U_i\Psi_iU_i^{\dag}, 
\label{gauge}
\end{equation}
where $g$ is the coupling constant and $U_i \equiv U_i(x^{\mu})$ is a $N_c \times N_c$
unitary matrix. In all earlier work on the transverse lattice
\cite{rev} $\Psi_i$ was in the fundamental representation.

The link variable can be written as 
\begin{equation}
  M_i(x^{\mu})=\exp\left( iagA_{i+1/2,\perp}(x^{\mu})\right), \label{M}
\end{equation}
where $A_{i,\perp} \equiv A_{i,2}$ is the transverse component of the gauge
potential at site $i$ and as $a \to 0$ we can formally expand Eq. \eqref{M} in
powers of $a$ as follows:
\begin{eqnarray}
M_i(x^{\mu})&=&1+iagA_{i+1/2,\perp}(x^{\mu})-\frac 12 \left(ag A_{i+1/2,\perp}
   (x^{\mu})\right)^2 +\ldots \nonumber \\
   &=&1+iagA_{i,\perp}(x^{\mu})+\frac {a^2}2 \left[ ig \partial_{\perp}
   A_{i,\perp}(x^{\mu})-g^2 \left(A_{i,\perp}
   (x^{\mu})\right)^2 \right]+O(a^3). 
\label{expandM}
\end{eqnarray}
 In the limit $a \to 0$, with the substitution of the expansion Eq.~\eqref{expandM} for
$M_i$, we expect everything to coincide with its counterpart in {\it
continuum} (2+1)--dimensional theory.

The discrete Lagrangian is then given by
\begin{eqnarray}
 {\cal L}&=&\tr \Bigl\{-\frac 14F_i^{\mu\nu}
  F_{i,\mu\nu} +\frac 1{2a^2g^2}\left(D_{\mu}M_i\right)
  \left(D^{\mu}M_i\right)^{\dag} \nonumber \\  
  &+& \bar{\Psi}_i i \gamma^{\mu}D_{\mu}\Psi_i+\frac i{2a}\bar{\Psi}_i
   \gamma^{\perp}(M_i\Psi_{i+1} M_i^{\dag}-M_{i-1}^{\dag}\Psi_{i-1}M_{i-1})
   \Bigr\},
    \label{lagrangian}
\end{eqnarray}
where the trace has been taken with respect to the color indices, $F_{i,\mu\nu}=
\partial_{\mu}A_{i,\nu}-\partial_{\nu}A_{i,\mu}+ig[A_{i,\mu},A_{i,\nu}]$, $\mu$, $\nu$ =
$\pm$ and $\gamma$'s are defined as follows  
\[ 
   \gamma^+ \equiv \frac{\gamma^0+\gamma^1}{\sqrt 2} \equiv 
   \frac{\sigma_2+i\sigma_1}{\sqrt 2}, \qquad
   \gamma^- \equiv \frac{\gamma^0-\gamma^1}{\sqrt 2} \equiv 
   \frac{\sigma_2-i\sigma_1}{\sqrt 2} , \qquad
   \gamma^{\perp} \equiv i\sigma_3, 
\]
and the covariant derivative $D_{\mu}$ is defined as
\begin{eqnarray}
D_{\mu}\Psi_i&=&\partial_{\mu}\Psi_i+ig[A_{i,\mu},\Psi_i], \label{covpsi}\\
   D_{\mu}M_i&=&\partial_{\mu}M_i+igA_{i,\mu}M_{i}-igM_iA_{i+1,\mu} \ 
   \stackrel{a \to 0}{\longrightarrow} \ 
   iag F_{\mu \perp},  \nonumber \\
   (D^{\mu}M_i)^{\dag}&=&\partial^{\mu}M_i^{\dag}-igM_{i}^{\dag}A_{i}^{\mu}
   +igA_{i+1}^{\mu}M_i^{\dag} \ 
   \stackrel{a \to 0}{\longrightarrow} \ 
   iag F^{\mu \perp}. \nonumber 
\end{eqnarray}
Thus, in the limit $a\to 0$ one finds, as expected,
\[  {\cal L} \ \stackrel{a \to 0}{\longrightarrow} \ 
     \tr \left(-\frac 14 F^{\a\beta}F_{\a\beta}
              +i\Psi i\gamma^{\a}D_{\a}\Psi\right),
\]              
where $\a$, $\beta=\pm,\perp$. Of course the form of this
Lagrangian is slightly different from that in Ref. \cite{rev} since the
fermions are in the adjoint representation. This  Lagrangian is
hermitian and invariant under the transformation in Eq.~\eqref{gauge} as one would expect.

The following Euler-Lagrange equations in the light cone gauge, $A_{i,-}=0$,
are constraint equations.
\begin{equation}
  \partial_-^2A_i^-\equiv gJ_i^+ ,\qquad
  \partial_- \chi_i=\frac 1{2\sqrt 2a}(M_i\psi_{i+1} M_i^{\dag}
  -M_{i-1}^{\dag}\psi_{i-1}M_{i-1}) \
   \stackrel{a \to 0}{\longrightarrow} \ \frac 1{\sqrt 2} D_{\perp} \psi, 
   \label{equations}
\end{equation}  
where   
\begin{eqnarray}
J_i^+ &\equiv &
  \frac i{2g^2a^2} (M_i\stackrel{\leftrightarrow}{\partial}_-
  M_i^{\dag}+M_{i-1}^{\dag}\stackrel{\leftrightarrow}{\partial}_-M_{i-1})
  +2\psi_i\psi_i \\ &\stackrel{a \to 0}{\longrightarrow} &\ 
  i[A_{\perp},\partial_-A_{\perp}]
    +\frac 1g \partial_-\partial_{\perp}A_{\perp} + 2\psi\psi, \\ 
  \Psi_i &\equiv& \frac {1}{2^{1/4}}
\left(\begin{array}{c}\psi_i \\ \chi_i \end{array}\right).
\end{eqnarray}
Since these equations only involve  the spatial derivative we can solve them for
$A_i^-$ and
$\chi_i$, respectively. Thus the dynamical field degrees of freedom are $M_i$,
$M_i^{\dag}$ and $\psi_i$. 

The first of the equations in Eq. \eqref{equations} gives a constraint on physical
states $|phys \ket$, since the zero mode of $J^+_i$ acting on any physical state
must vanish, 
\begin{equation}
  \stackrel{0}{J^+_i}|phys \ket=\int dx^- J^+_i(x^{\mu})|phys \ket=0 \quad 
  {\rm for} \ {\rm any} \ i. 
\label{constraint}
\end{equation}
The physical states must be color singlet at {\it each} site.
 
It is straightforward to derive $P^{\pm}\equiv \int dx^- T^{+\pm}$, where
$T^{\mu\nu}$ is the stress-energy tensor. We have 
\begin{eqnarray}
 P^+&=&a\sum_{i=1}^{N_{sites}}\int dx^- \tr\left( \frac 1{a^2g^2}
    \partial_-M_i^{\dag} \partial_- M_i +i\psi_i\partial_-\psi_i\right) 
    \label{pplus} \\ 
    &\stackrel{a \to 0}{\longrightarrow}& \int dx^- dx^{\perp} \tr\left( 
    (\partial_- A_{\perp})^2 +i\psi\partial_-\psi\right)
    \label{contpplus}, 
\end{eqnarray}
and
\begin{eqnarray}
 P^-&=&a\sum_{i=1}^{N_{sites}}\int dx^- \tr\left[ \frac 12(
     \partial_-A_i^-)^2 +i\chi_i\partial_-\chi_i\right] \nonumber \\    
     &=&a\sum_{i=1}^{N_{sites}}\int dx^- \tr\Bigl[  -\frac {g^2}2 J^+_i
       \frac 1{\partial_-^2}J^+_i\nonumber \\ 
       &&-\frac i{8a^2}(M_i\psi_{i+1} M_i^{\dag}
     -M_{i-1}^{\dag}\psi_{i-1}M_{i-1})\frac 1{\partial_-}(M_i\psi_{i+1} 
     M_i^{\dag}-M_{i-1}^{\dag}\psi_{i-1}M_{i-1})\Bigr] \label{pminus} \\
     &\stackrel{a \to 0}{\longrightarrow}&\int dx^- dx^{\perp} 
     \tr\Bigl[ -\frac {g^2}2 J^+\frac 1{\partial_-^2}J^+
       -\frac i{2}D_{\perp}\psi \frac 1{\partial_-}D_{\perp}\psi 
     \Bigr]. \label{contpminus}
\end{eqnarray} 
When one quantizes the dynamical fields, unitarity of $M_i$ is lost and $M_i$ becomes an $N_c
\times N_c$ imaginary matrix
\cite{dv,rev}. Some have suggested the addition of an
effective potential $V(M_i)$ to force $M_i$ to be a unitary matrix in the
limit $a \to 0$ \cite{bardeen,BPR80,rev}. We will approach this issue
using supersymmetry.

Having
linearized $M_i$, we can expand $M_i$ and $\psi_i$ in their Fourier modes as follows; at
$x^+=0$
\begin{eqnarray}
 M_{i,rs}(x^-)&=&\frac {ag}{\sqrt{2\pi}}\int_0^{\infty} \frac{dk^+}{\sqrt{k^+}}
   (d_{i,rs}(k^+)\e^{-ik^+x^-}+a_{i,sr}^{\dag}(k^+)\e^{ik^+x^-}), \label{m} \\ 
  \psi_{i,rs}(x^-)&=&\frac 1{2\sqrt{\pi}}\int_0^{\infty}  dk^+
   (b_{i,rs}(k^+)\e^{-ik^+x^-}+b_{i,sr}^{\dag}(k^+)\e^{ik^+x^-}), \label{psi}
\end{eqnarray}
where $r,s$ indicate the color indices, $a_{i,sr}^{\dag}(k^+)$ creates a link variable
with momentum $k^+$ which carries color $r$ at site $i$ to $s$ at site $(i+1)$,
$d^{\dag}_{i,sr}(k^+)$ creates a link with $k^+$ which carries color $r$ at site
$(i+1)$ to $s$ at site $i$ and $b^{\dag}_{i,sr}$ creates a fermion at the $i$-site
which carries color $r$ to $s$. Quantizing at $x^+=0$ we have 
\begin{eqnarray}
&&[M_{i,rs}(x^-),\pi_{M_j,pq}(y^-)]=   
[M_{i,rs}^{\dag}(x^-),\pi_{M^{\dag}_j,pq}(y^-)]\nonumber\\
   &&=\{ \psi_{i,rs}(x^-),\pi_{\psi_j,pq}(y^-)\}=\frac i2 \delta(x^--y^-)
   \frac {\delta_{ij}}a \delta_{rp}\delta_{sq}.
\end{eqnarray}
Note that we divided $\delta_{ij}$ by $a$ because $\delta_{ij}/a \to
\delta(x^{\perp}-y^{\perp}) $as $a \to 0$.  The conjugate momentum are 
\[
   \pi_{M_i} 
    =\frac 1{2a^2g^2} \partial_- M_i^{\dag}, \quad
   \pi_{M_i^{\dag}}   
   =\frac 1{2a^2g^2} \partial_- M_i, \quad
   \pi_{\psi_i} 
    =i\psi_i.
\]
Thus we must have 

\begin{eqnarray}
[M_{i,rs}(x^-),\partial_{-y}M^{\dag}_{j,pq}(y^-)]&=&
   [M_{i,rs}^{\dag}(x^-),\partial_{-y}M_{j,pq}(y^-)]=
   ia^2g^2\delta(x^--y^-)\frac {\delta_{ij}}a\delta_{rp}\delta_{sq}, \nonumber \\
   \{ \psi_{i,rs}(x^-),\psi_{j,pq}(y^-)\}&=&\frac 12 \delta(x^--y^-)
   \frac {\delta_{ij}}a \delta_{rp}\delta_{sq}.
\end{eqnarray}
Then, one can easily see that these commutation relations are satisfied when $a$'s,
$d$'s and
$b$'s satisfy the following:

\begin{equation}
 [a_{i,rs}(k^+),a^{\dag}_{j,pq}(p^+)]=[d_{i,rs}(k^+),d^{\dag}_{j,pq}(p^+)]
   =\{ b_{i,rs}(k^+),b^{\dag}_{j,pq}(p^+)\}=\delta(k^+-p^+)\frac {\delta_{ij}}a     \delta_{rp}\delta_{sq}, \label{commutations}
\end{equation}
with others all being zero. 
Physical states can be generated by acting on the Fock
vacuum $|0 \ket$ with these $a^{\dag}$'s, $d^{\dag}$ and $b^{\dag}$'s in such a manner
that the constraint Eq.~\eqref{constraint} is satisfied.  

\begin{figure}[ht]
\begin{tabular}{cc}
\psfig{figure=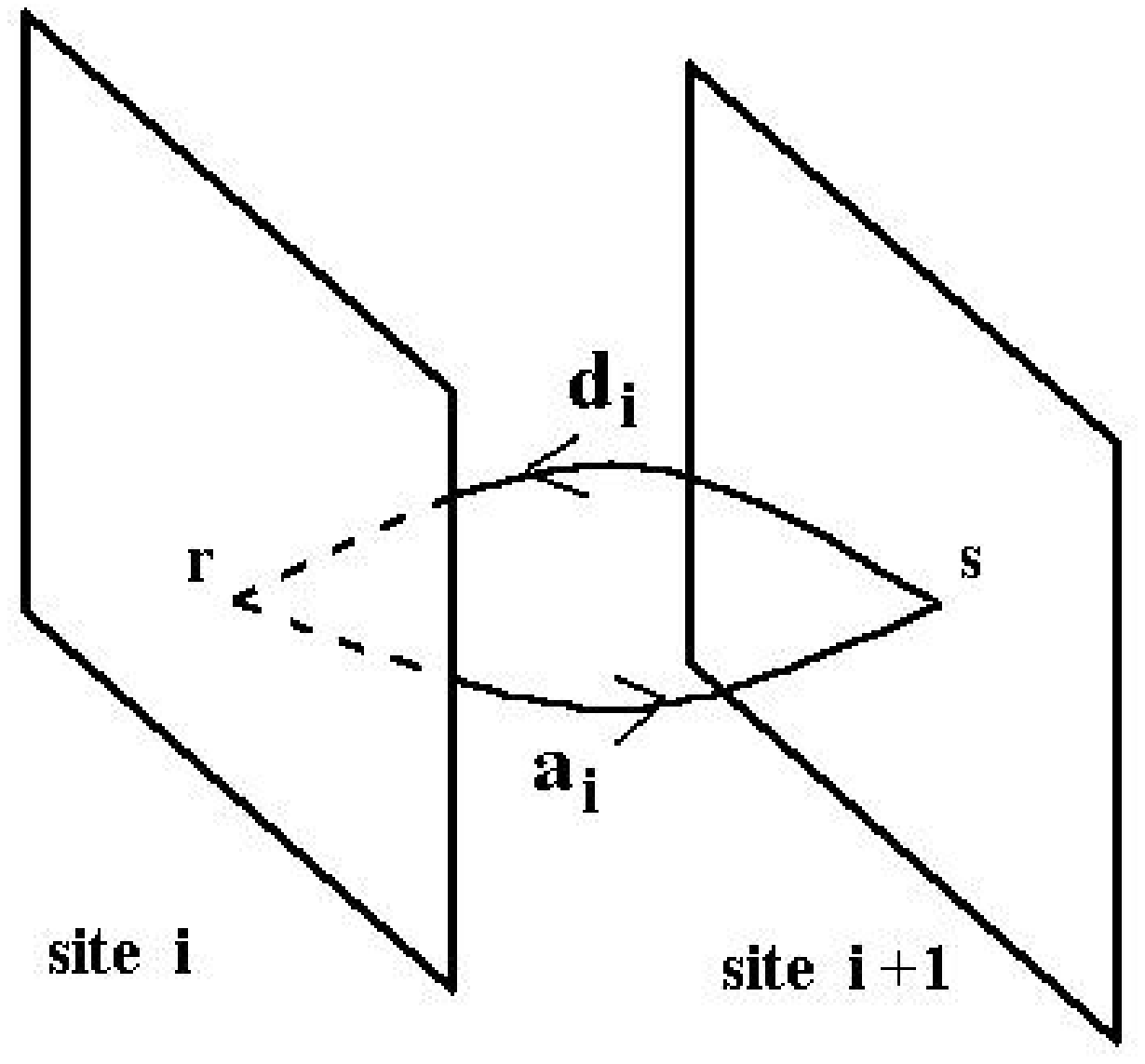,width=7.5cm,angle=0}&
\psfig{figure=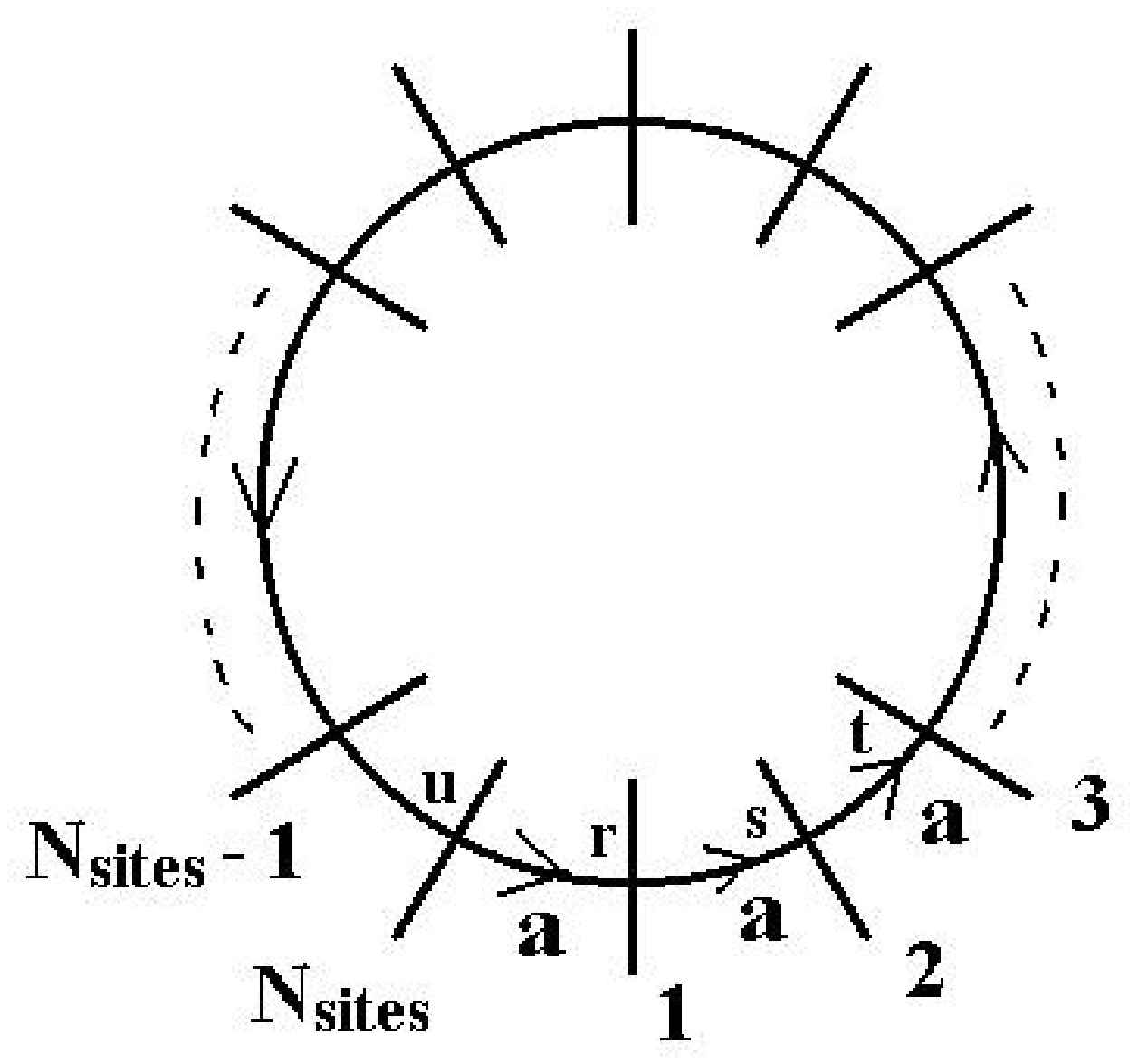,width=7.5cm,angle=0}\\
(a) & (b)
\end{tabular}
\caption{(a)The color charge for the state $|phys \ 1\ket \equiv
d^{\dag}_{i,rs}(k_1^+)a^{\dag}_{i,sr}(k^+_2)|0\ket$. The planes represent the color
space. $a_i$ carries color $r$ at site $i$ to $s$ at site $i+1$ and $d_i$ carries it
back to $r$ at site $i$. (b) The color for the state $|phys \ 2\ket \equiv
a^{\dag}_{i+N_{sites}-1,ru}(k^+_{N_{sites}})\cdots a^{\dag}_{i+1,ts}(k^+_2)\cdot 
a^{\dag}_{i,sr}(k^+_1)|0\ket$. The lines which intersect a circle
represent the color planes at sites. The color goes all the way around the transverse
lattice.}
\label{da}
\end{figure}

Let us complete this section
by discussing the physical constraint \eqref{constraint} in more detail.
The states are all constructed in the large--$N_c$ limit, and therefore we need only
consider single--trace states.
 In order for a state to be color
singlet at each site, {\it each color index has to be contracted at the same site}. As
an example consider a state represented by
$|phys \ 1\ket \equiv d^{\dag}_{i,rs}(k_1^+)a^{\dag}_{i,sr}(k^+_2)|0\ket$.  For this state
the color $r$ at site $i$ is carried by $a^{\dag}_i$ to $s$ at site $(i+1)$ and then
brought back by $d^{\dag}_i$ to $r$ at site $i$. The color $r$ is contracted at site $i$
{\it only} and the color $s$ at site $(i+1)$ {\it only}. Therefore, this is a physical
state satisfying Eq. \eqref{constraint}. A picture to visualize
this case is shown in Fig. \ref{da}a.
One also needs to be careful with operator ordering. One can show that the state
$ d^{\dag}_{i,rs}(k^+_1)a^{\dag}_{i,st}(k^+_2)b^{\dag}_{i,tr}(k^+_3)|0\ket$ is
physical, while the state
$b^{\dag}_{i,rs}(k^+_1)a^{\dag}_{i,st}(k^+_2)d^{\dag}_{i,tr}(k^+_3)|0\ket$ is
unphysical. 

We should, however, note that a true physical state be summed over all the  transverse
sites since we have discrete translational symmetry in the transverse  direction. That
is, for example, the states 
$d^{\dag}_{1,rs}(k_1^+)a^{\dag}_{1,sr}(k^+_2)|0\ket$ and 
$d^{\dag}_{2,rs}(k_1^+)a^{\dag}_{2,sr}(k^+_2)|0\ket$ are the same up to a phase 
factor given by $\exp(iP^{\perp}a)$. We set the phase factor to one since we take
physical state to have
$P^{\perp}=0$.   The physical state $|phys \ 1\ket$ is in fact 
$\sum_{i=1}^{N_{sites}}d^{\dag}_{i,rs}(k_1^+)a^{\dag}_{i,sr}(k^+_2)|0\ket$  with the
appropriate normalization constant. From a computational point of view this is a
great simplification because we can drop the site index i from the representation.

Periodic conditions on the
fields, allow for physical states of the form $|phys \ 2\ket \equiv \sum_i 
a^{\dag}_{i+N_{sites}-1,ru}(k^+_{N_{sites}})\cdots a^{\dag}_{i+1,ts}(k^+_2)\cdot 
a^{\dag}_{i,sr}(k^+_1) |0\ket$. The color for this state is
carried around the transverse lattice, as shown in Fig.~\ref{da}b.  We will refer to these
states as cyclic states. The states where the color flux does not go all the way around the
transverse lattice we will refer to as non-cyclic states. We characterize  states by what we
call the winding number defined by $W=n/N_{sites}$, where
$n\equiv \sum_i(a_i^{\dag}a_i-d^{\dag}_id_i)$. Using the Eguchi--Kawai\cite{egk83} reduction
which applies in the large--$N_c$ limit we can always take $N_{sites} = 1$. The winding
number simply  gives us the excess number 
of $a^{\dag}$ over $d^{\dag}$ in a state. We use the winding number to classify
states since the winding number is a  good quantum number commuting with
$P^-_{SDLCQ}$. In the language of the winding
number the non-cyclic states  are those states with $W=0$ and cyclic
states have non-zero $W$.

It is straight forward to show that $| phys \ket $ satisfies Eq.~\eqref{constraint} but
$|unphys \ket$  does not using
\begin{eqnarray}
(\stackrel{0}{J^+_i})_{pq} &=& \int dk^+ a^{\dag}_{i,rp}(k^+)a_{i,rq}(k^+)
    -d_{i,pr}(k^+)d^{\dag}_{i,qr}(k^+)-a_{i-1,pr}(k^+)a^{\dag}_{i-1,qr}(k^+) 
\nonumber \\  &&
    +d^{\dag}_{i-1,rp}(k^+)d_{i-1,rq}(k^+)+b_{i,pr}(k^+)b^{\dag}_{i,qr}(k^+)
    +b^{\dag}_{i,rp}(k^+)b_{i,rq}(k^+).
\end{eqnarray}
Diagrammatically, one can say that at every point in color space at any site one has to
have either no lines or {\it two} lines, one of which goes into and the other of which
comes out of the point, so that the color indices are contracted at the same site.

\section{SDLCQ of the Transverse Lattice Model}

The transverse lattice formulation of ${\cal N}=1$ SYM theory in 2+1 dimension presented
in the previous section has several undesirable features. The supersymmetric structure of
the theory is completely hidden and the resulting Hamiltonian is $6^{th}$ order in the
fields. From the numerical point of view a $6^{th}$ order interaction makes the theory
considerably more difficult to solve. Also the underlying (2+1)--dimensional
supersymmetric Hamiltonian is only $4^{th}$ order making this discrete formulation of the 
theory very different than the underlying theory. There can, of course, be many discrete
formulations that correspond to the same continuum theory and it is therefore desirable to
search for a better one. In the spirt of SDLCQ we will attempt a discrete formulation based
on the underlying super algebra of this theory,
  
\begin{equation}
 \{Q^{\pm},Q^\pm\}=2\sqrt 2 P^\pm, \quad    \{Q^+,Q^-\}=2P^{\perp}. 
\end{equation}

In this effort there are some fundamental limits to how far one can go. As we discussed in
the previous section the physical states of this theory must conserve color at every point
on the transverse lattice. Experience with other supersymmetric theories indicates
that each term in
$Q^+$ has to be either the product of {\it one}
$M_i$ and {\it one}
$\psi_i$ or of {\it one} $M_i^{\dag}$ and {\it one} $\psi_i$ therefore $Q^+$ is {\it
unphysical}, by which we mean that $Q^+$ transforms a physical state into an unphysical
one, so that $\bra phys|Q^+|phys \ket=0$. While this is not a theorem, it seems very
difficult to have any other structure since in light cone quantization $P^+$ is a
kinematic operator and therefore independent of the coupling. There appears to be  {\it
no} way to make a physical
$P^+$ from
$Q^+$. We will use
$P^+$ as given in Eq.
\eqref{pplus} in what follows. Similarly, we are not able to generally construct
$P^{\perp}$ from
$Q^+$ and $Q^-$. In fact $P^{\perp}$ is 
unphysical in our formalism, leading to $\bra phys|P^{\perp}|phys \ket=0$. Formally we
will work in the frame where total $P^\perp$ is zero, so it would appear consistent
with this result. However, $P^\perp=0$ was a choice and a non-zero value is equally
valid and not consistent with the matrix element. 

Despite these difficulties we find a physical $Q^-$
which gives us $P^-_{SDLCQ}\stackrel{a\to 0}{\longrightarrow} P^-_{cont}$. The
expression for $Q^-$ and $P^-_{SDLCQ}$ are, respectively, 

\begin{eqnarray}
 Q^- &=& 2^{3/4}g\cdot a \sum_{i=1}^{N_{sites}}\int dx^- \tr (J_i^+
        \partial^{-1}_- \psi_i) \label{qminus} \\
      &\stackrel{a \to 0}{\longrightarrow} &
2^{3/4}\int dx^-dx^{\perp}
        \tr \left[\partial_{\perp}A^{\perp}\psi +g\left( 
        i[A^{\perp},\partial_-A^{\perp}]+ 2\psi\psi\right)
        \partial^{-1}_- \psi\right],  \nonumber  
\end{eqnarray}
\begin{eqnarray}
P^-_{SDLCQ}&\equiv& \frac{\{ Q^-,Q^-\}}{2\sqrt 2} \nonumber \\
  &=&a\sum_i \int dx^- \tr [ -\frac {g^2}2 J_i^+\frac 1
  {\partial_-^2}J_i^+ -\frac i{2a^2} (\psi_{i+1}M^{\dag}_i-M_i^{\dag}\psi_i)
  \partial^{-1}(M_i\psi_{i+1}-\psi_iM_i)] \label{qminusqminus} \nonumber\\
  &\stackrel{a\to 0}{\longrightarrow}&  \int dx^- dx^{\perp} 
     \tr\Bigl[ -\frac {g^2}2 J^+\frac 1{\partial_-^2}J^+
       -\frac i{2}D_{\perp}\psi \frac 1{\partial_-}D_{\perp}\psi 
     \Bigr]\equiv 2\sqrt 2 P^-_{cont}. 
\end{eqnarray}
Notice that this Hamiltonian is only $4^{th}$ order in the fields. 
Furthermore, one can check that this $Q^-$ commutes with $P^+$ obtained from ${\cal
L}$; $[Q^-,P^+]=0$. Thus, it follows that,
\begin{equation}
  \bra phys|[Q^-,M^2]|phys\ket =\bra phys|[Q^-,2P^+P^-_{SDLCQ}]|phys\ket =0
  \label{supersymmetry}
\end{equation}
in our SDLCQ formalism, where $M^2 \equiv 2P^+P^-_{SDLCQ}-(P^{\perp})^2$. The fact that the
Hamiltonian is the square of a supercharge will guarantee the usual supersymmetric degeneracy
of the massive spectrum, and our numerical solutions will substantiate this. Unfortunately one
needs a $Q^+$ to guarantee the degeneracy of the massless bound states.

The expression for $Q^-$ is  

\begin{eqnarray}
 :Q^-:&=& \frac{i2^{-1/4} ag}{\sqrt{\pi}}\sum_i \int dk_1 \ dk_2 \ dk_3 
         \delta(k_1+k_2-k_3) \Bigl[ \nonumber \\
      && \frac{k_2-k_1}{k_3\sqrt{k_1k_2}}(-b_i^{\dag}d_ia_i
        +d_i^{\dag}a_i^{\dag}b_i-b^{\dag}_ia_{i-1}d_{i-1}
        +a^{\dag}_{i-1}d_{i-1}^{\dag}b_i) \nonumber \\
      && +\frac{k_2+k_3}{k_1\sqrt{k_2k_3}}(-d_i^{\dag}b_id_i
        +b_i^{\dag}d_i^{\dag}d_i-a^{\dag}_{i-1}b_{i}a_{i-1}
        +b^{\dag}_{i  }a_{i-1}^{\dag}a_{i-1}) \nonumber \\     
      && +\frac{k_3+k_1}{k_2\sqrt{k_3k_1}}( a_i^{\dag}a_ib_i
        -a_i^{\dag}b_i^{\dag}a_i+d^{\dag}_{i-1}d_{i-1}b_{i  }
        -d^{\dag}_{i-1}b_{i  }^{\dag}d_{i-1}) \nonumber \\     
      && +\left( \frac 1{k_1}+\frac 1{k_2} -\frac 1{k_3}\right)
       (b^{\dag}_ib^{\dag}_ib_i+b^{\dag}_ib_ib_i) \Bigr], \label{qminusinabd}
\end{eqnarray}
where  $k^+ \equiv k$, $a^{\dag}a\equiv {\rm Tr}(a^{\dag}(k_1)a(k_2))$,
$a^{\dag}aa\equiv {\rm Tr}(a^{\dag}(k_3)a(k_1)a(k_2))$,
$a^{\dag}a^{\dag}a\equiv {\rm Tr}(a^{\dag}(k_1)a^{\dag}(k_2)a(k_3))$. Notice that
from this explicit expression for $Q^-$ it is clear that cyclic states do not get
mixed with non-cyclic states under $Q^-$, as advertised at the end of the second
section.  Notice also that the winding number introduced in the last section
evidently  commutes with $Q^-$ and, thus, with $P^-_{SDLCQ}$.

Now we are in a position to solve the 
eigenvalue problem $ 2P^+P^-_{SDLCQ}|phys\ket=m^2|phys\ket$.   We impose the periodicity
condition on $M_i$, $M^{\dag}_i$ and $\psi_i$ in the $x^-$ direction giving a discrete spectrum
for
$k^+$:
\[ 
k^+=\frac {\pi}L n \quad (n=1,2,\ldots.), \qquad \int_0^{\infty} dk^+ 
   \rightarrow \frac {\pi}L \sum_{n=1}^{\infty}.  
\]
We impose a cut-off on the total longitudinal momentum $P^+$ i.e. $P^+=\pi K/L$, where
$K$ is an integer also known as the `harmonic resolution', which indicates the coarseness
of our numerical results.  For a
fixed $P^+$ i.e. a fixed $K$, the number of partons in a state is limited up to the
maximum, that is $K$, so that the total number of Fock states is {\it finite}, and, therefore,
we have reduced the infinite dimensional eigenvalue problem to a finite dimensional one. 
We should note here that since the matrix $\bra phys|P_{SDLCQ}^-|phys \ket$ 
to be diagonalized does not depend on $N_{sites}$, the resulting spectrum does not 
depend on $N_{sites}$, either. This means there is no need to keep the site index 
of operators even in numerical calculations; the sum over all the sites is 
implicitly understood and when one needs to restore the site indices for some reason, 
one should do so in such a way that physical constraint \eqref{constraint} is satisfied.
Henceforth
we will suppress the sum and the site indices,  unless otherwise noted.

For this initial study of the transverse lattice  we only consider resolution 
up to $K=8$ for $W=0,1$ states and up to $K=W+5$ and $K=W+4$ for states with 
$|W|=2,3$ and $|W|=4,5$, respectively.  We were able to handle these 
calculations with our SDLCQ Mathematica code. 
In the following two sections we will give the numerical results for the
cyclic $(W\ne 0)$ states and non-cyclic $(W = 0)$ states separately.

\section{Numerical Results for the Cyclic $(W\ne 0)$ States}

For the cyclic states, it is easy to see that $K\ge |W|$. In fact if
$K=|W|$, only two states are possible and both are bosonic they are
$\tr( a^{\dag}_{i+N_{sites}-1}\cdots a^{\dag}_{i+1} a^{\dag}_{i})|0\ket $ and
$\tr( d^{\dag}_id^{\dag}_{i+1}\cdots d^{\dag}_{i+N_{sites}-1})|0\ket$, 
Therefore we will focus on $K > |W|$.  Since there is an exact $Z_2$
symmetry between positive $W$ and negative $W$, it suffices to consider the 
case where $W$ is positive.
\begin{table}[t]
\centerline{
\begin{tabular}{|c|ccccccc|}
\hline
 K--W&1 & 2& 3& 4& 5& 6& 7 \\
\hline
\multicolumn{8}{|c|}{ $W$  \quad massive fermion or boson states}\\
\hline
1 & 0& 1&  5&  18&  62&  208&706 \\
2 & 0& 2& 10&  38& 138&  492& \\
3 & 0& 3& 17&  68& 268& 1023& \\
4 & 0& 4& 24& 110& 470&     & \\
5 & 0& 5& 33& 166& 770&     & \\
\hline
\multicolumn{8}{|c|}{ massless boson states}\\
\hline
1 & 0& 1&  1&  3&  3& 8  & 8 \\
2 & 1& 2&  2&  5&  5& 12 & \\
3 & 1& 2&  2&  5&  5& 15 & \\
4 & 1& 2&  2&  6&  6&    & \\
5 & 1& 2&  2&  6&  6&    & \\
\hline
\multicolumn{8}{|c|}{ massless fermion states}\\
\hline
1 & 1& 1& 2&  2&  4&  4& 9 \\
2 & 1& 1& 2&  2&  5&  5& \\
3 & 1& 1& 2&  2&  5&  5& \\
4 & 1& 1& 2&  2&  6&   & \\
5 & 1& 1& 2&  2&  6&   & \\
\hline
\end{tabular}}
\caption{Number of massive and massless cyclic eigenstates.}
\label{cirmass}
\end{table}
Table~\ref{cirmass} shows the number of eigenstates with different $K$
and $W$ for various types of states. Since the spectrum starts at  
$K = W$, it is natural to take $K-W$ 
as the independent variable. Therefore we tabulate the number of eigenstates with $W$ and
$K-W$ rather than $W$ and $K$ and we
plotted $m^2$ as a function $1/(K-W)$
 rather than in $1/K$ 

The massive degenerate fermion and boson states are related by $Q^-$
$|F\ket\equiv |B\ket$.  The same is not true of massless states. There is no
direct connection through
$Q^-$ between massless fermionic states and massless bosonic states, leading to a
supersymmetry breaking for massless states. 
Nevertheless, Table \ref{cirmass} shows that we
have the exact supersymmetry for massless states when $K-|W|=2n-1$ for           
$n=2,\ldots$. The boson state with $W=1$ is anomalous since $\tr(a^{\dag})=0$ in our
formulation.

Also notice that there is a jump in the number of massless states
with every increment by two in $K$.  This seems to be the 
case because we need to increase $K$ by two to allow for the addition of an operator 
like $d_i^{\dag}(1)a_i^{\dag}(1)$, so as to make a new physical massless state. 
The requirement that we add a pair of bosons relates back to the Gauss's
law constraint. We see here that two bosons are behaving as a single boson. 
This is additional evidence that Gauss's law and supersymmetry are
working together to restrict the number of effective boson degrees of freedom.
It is particularly reassuring to see this effect in the massless bound states
since it is in this sector where breaking of the supersymmetric spectrum occurs.
We also notice some other interesting properties of our  massless states.
We find that the Fock states that occur in
bosonic massless states have {\it no} fermionic operators, whereas the Fock
states that occur in fermionic massless states have only {\it one} fermionic
operator, which seems to explain the relative shift between the number of massless fermion and
bosons.

\begin{figure}[t!]
\begin{tabular}{cc}
\psfig{figure=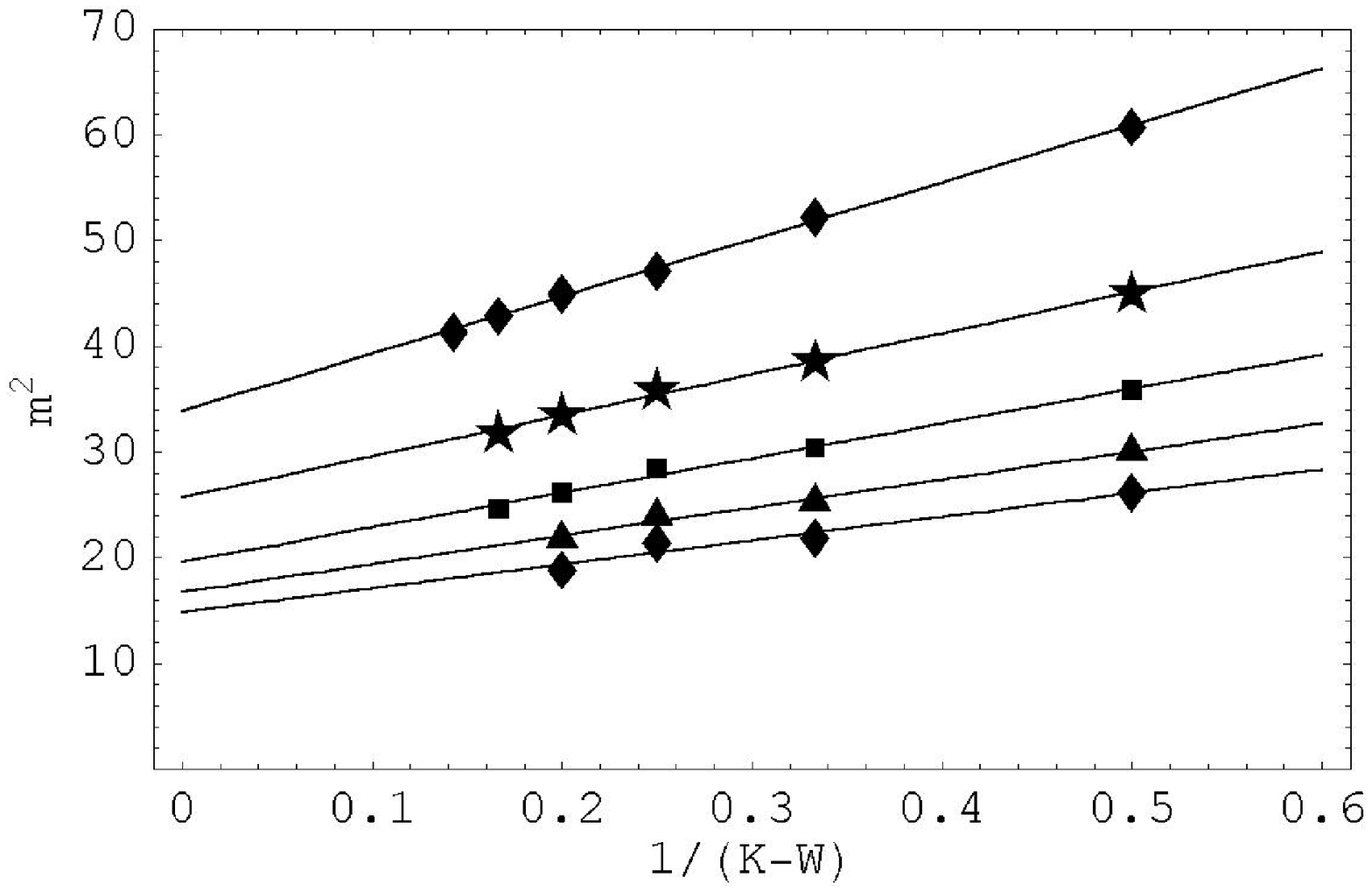,width=7.5cm,angle=0}&
\psfig{figure=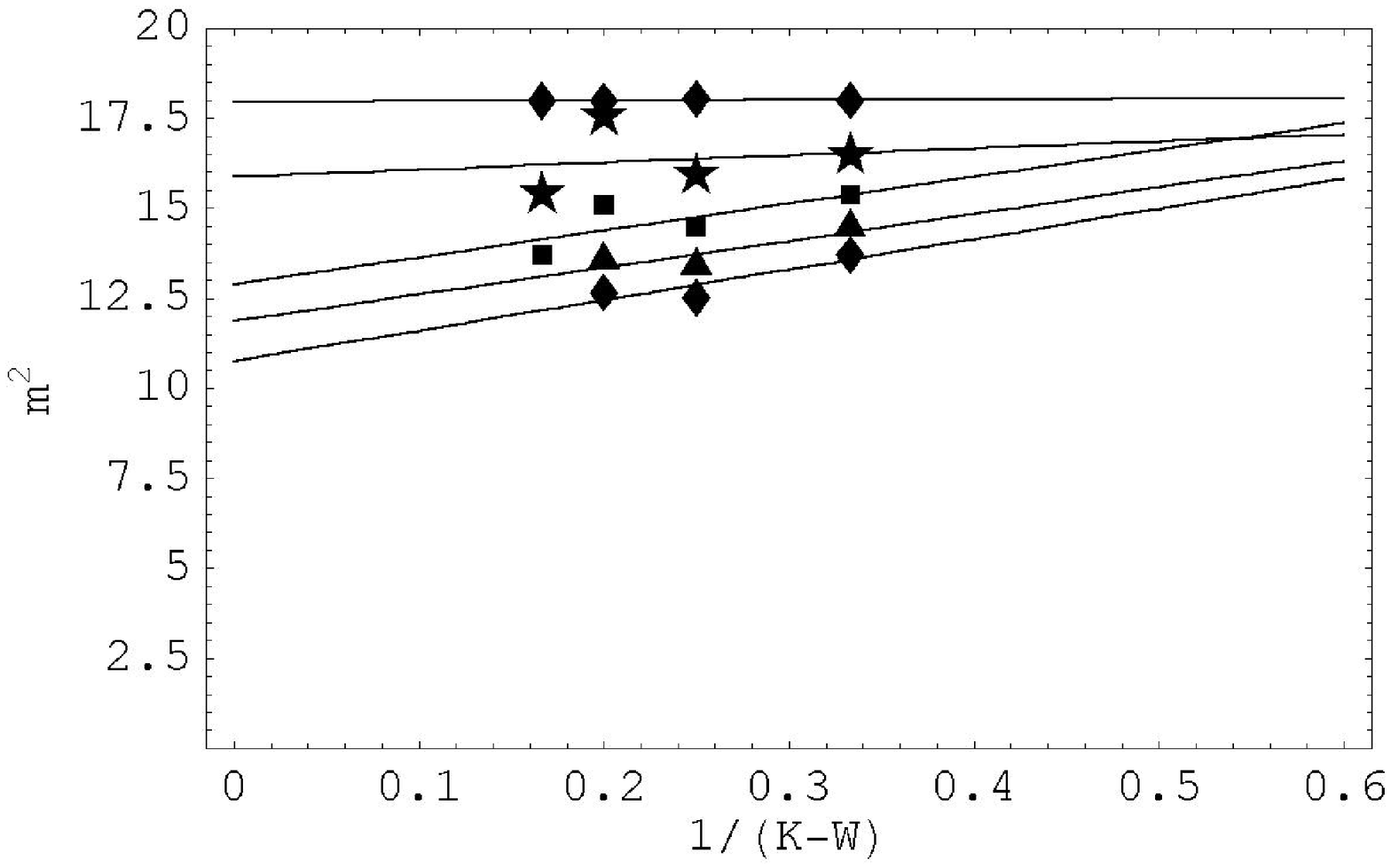,width=7.5cm,angle=0}\\
(a)&(b)
\end{tabular}
\caption{Plots of $m^2$ in units 
of $\frac{N_cg^2}{\pi a}$ of low energy cyclic states versus  $1/(K-W)$ with a linear fit 
for W=1(top diamond), 2(star), 3(square), 4(triangle), 5(bottom diamond) (a) state A and, (b) 
state B. }
\label{fit:cyclic}
\end{figure}

In Fig.~\ref{fit:cyclic}(a) and (b) we give plots of $m^2$ for two low--energy
states  as a function of $1/(K-W)$ and extract $m^2_\infty$ as a $K \to \infty$ limit of the
linear fit. We identify an  energy eigenstate with different $K$'s according to dominant Fock
states. Looking at both bosonic  and fermionic counterpart also helps
distinguish states. We present two states we could easily identify. 
For the state in
(a) the dominate fock component has the form
$b^{\dag}(n)a^{\dag}(1)\cdots a^{\dag}(1) b^{\dag}(1)+b^{\dag}(1)a^{\dag}(1)\cdots
a^{\dag}(1)b^{\dag}(n)$ while the state in (b) has the dominate component
$b^{\dag}(n)a^{\dag}(1)\cdots a^{\dag}(1) b^{\dag}(1)-b^{\dag}(1)a^{\dag}(1)\cdots
a^{\dag}(1)b^{\dag}(n)$. 

\begin{figure}[t!]
\begin{center}
\begin{tabular}{c}
\psfig{figure=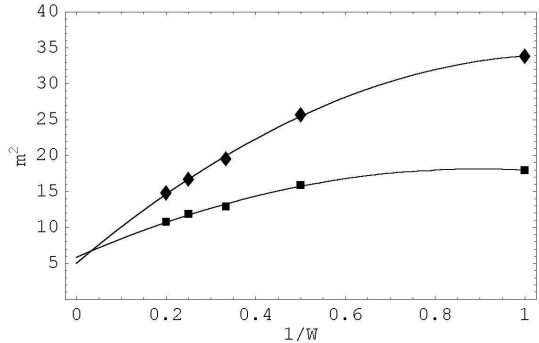,width=7.5cm,angle=0}
\end{tabular}
\end{center}
\caption{Plots of $K \to \infty$ limit of $m^2$ in units 
of $\frac{N_cg^2}{\pi a}$ of low energy cyclic states versus $1/W$  with a
quadratic fit to the data. The diamond correspond state A box and squares correspond to the state
B in Fig.~\ref{fit:cyclic}}
\label{fit:cyclic2}
\end{figure} 

In Fig.~\ref{fit:cyclic2} we present $m^2_\infty$, obtained in Fig.\ref{fit:cyclic}(a) and
(b), as a function of $1/W$ and  show a quadratic fit to the
data.  $m^2$ is fit very well with a quadratic fit in $1/W$
In 2+1 dimensions the Hamiltonian has the form $P^-=a + b k_\perp +c k_\perp^2$. With periodic
boundary conditions $k_\perp=n_\perp \pi/L$ and for cyclical states $L=aW$. Thus we expect that
$m^2$ is a function of the winding number to be of the form $m^2=A +B/W +C/W^2$, as we find.

This behavior is consistent with the unique properties of SYM theories that we
have seen in previous SDLCQ calculations
\cite{Antonuccio:1998kz,Antonuccio:1998jg}. We have seen that as we increase $K$ we discover new
lower mass bound states. Most of the partons in these long states appear to gluons.
Supersymmetric theories like to have light states with long strings of gluons. We call these
states with long strings of gluons, stringy states. In the full SDLCQ solution of the
${\cal N}=1$ SYM theory in 1+1 dimensions we found that these stringy states have an
accumulation point at zero mass. In the full SDLCQ calculation of
${\cal N}=1$ SYM theory in 2+1 dimensions we have seen stringy states as well, but we have not
seen evidence for an accumulation point. We will need to go to higher resolution to make a
similar analysis for this theory.

\section{Numerical Results for the Non-Cyclic $(W=0)$ States}

Let us now discuss numerical results for the non-cyclic states.  Table~\ref{masses_nc} shows
the number of mass eigenstates of massive bosons or fermions, massless bosons, and massless
fermions with different $K$.
\begin{table}
\centerline{
\begin{tabular}{|ccccccc|}
\hline
$K=$&3 & 4& 5& 6& 7& 8 \\
\hline
\multicolumn{7}{|c|}{  massive fermion or boson states}\\
\hline
&2&  6& 22&  72& 238& 792  \\
\hline
\multicolumn{7}{|c|}{  massless boson states}\\
\hline
&1& 3&  3&  7 &  7&   17 \\
\hline
\multicolumn{7}{|c|}{  massless fermion states}\\
\hline
&1&  1&  3& 3 &  7 &   7 \\
\hline
\end{tabular}}
\caption{ Number of massive and massless non-cyclic eigenstates}
\label{masses_nc}
\end{table}
From the table we see once again that there are some differences in the number
of the massless bosonic and fermionic states and the same dependence on $K$ that we
saw for the cyclic states. The reason for this behavior is  the same as in the
case of the cyclic states. 
\begin{figure}[t!]
\begin{center}
\begin{tabular}{c}
\psfig{figure=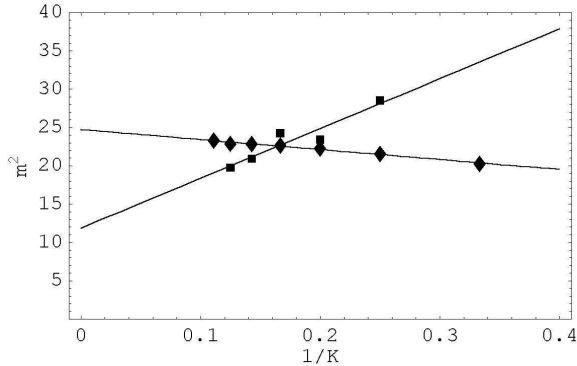,width=8cm}
\end{tabular}
\end{center}
\caption{Plots of $m^2$ of low--energy non-cyclic states against $1/K$ with 
a linear fit in units 
of $\frac{N_cg^2}{\pi a}$.}
\label{states}
\end{figure}
In Fig.~\ref{states} we show two states whose boson states with a large two partons component.
These states appear at the lowest  resolution and are the easiest to follow and identify as a
function of the resolution $K$. The boson bound state denoted by diamonds is composed primarily
of two fermions,
$b^{\dag}b^{\dag}$, while the boson bound state denoted by squares is composed primarily of two
bosons, $d^{\dag}a^{\dag}$. Again, we see stringy states which appear as we  go to higher $K$
with more partons in their dominate Fock state component.

We were able go up to $K=8$ without making any approximations to the Fock basis, so some of
our bound states contained as many as eight partons.  However, for $K=9$ we have truncated the
number of partons at 6.  We were able to justify this approximation  at $K=9$
for this state by comparing the truncated results with the exact result at $K=8$.
However we were not able to make this approximation for the state denoted by squares.

\section{Discussion}
We have presented a formulation of ${\cal N}=(1,1)$ SYM in 2+1 dimensions where the transverse
dimension is discretized on a spatial lattice while the longitudinal dimension $x^-$
is discretized on a momentum lattice. Both $x^-$ and $x^\perp$ are compact.  We are able to
retain some of the technology of SDLCQ, since this numerical approximation  retains one exact
supersymmetry. In particular we are able to write the Hamiltonian as the square of a
supercharge. Thus there is sufficient supersymmetry in this formulation to ensure that
divergences that appear in this theory are automatically canceled. Furthermore we show that
this formulation leads to a fundamentally different and simpler discrete Hamiltonian than the
standard Lagrangian approach to the transverse lattice. Since we only have one exact 
supersymmetry, only the massive fermion and boson bound states in
our solution are exactly degenerate. We need an additional supersymmetry to require that the
numbers of massless bosons and massless fermions be the same.

As in all transverse lattice approaches, the transverse gauge field is replaced by a
complex unitary field, and transverse gauge invariance is maintained. When this complex unitary
field is quantized as a general complex linear field, the number of degrees of freedom in the
transverse gauge field is improperly represented. In a conventional transverse lattice
calculation one tries to dynamically enforce the proper number of degrees of freedom by adding
a potential that is minimized by the unitarity constraint. We conjecture that this is not
necessary here.  Gauss's law requires that color be conserved at every transverse
lattice site. This greatly restrict the allowed boson Fock states that can be part of the
physical set of basis states and plays an important role in the structure of all bound states.
We  assert that the combination of the Gauss's law constraint and the one exact supersymmetric
are sufficient to approximately enforce the full supersymmetry.

To further support this conjecture we note that
in the massless spectrum the number of states changes when we change the resolution by two units
indicating that it effectively requires two partons to represent one true degree of freedom. We
view solving the theory as a unitary transformation from the constrained basis to a basis free
of constraints and very nearly fully supersymmetric. 

We should note that this conjecture can not be general  since we know of one supersymmetric
theory in 1+1 dimensions where the degrees of freedom at the parton level are all
fermions \cite{kut93}. In this model one has to fix the coupling to be a particular value for
this miracle to occur. Generally in a supersymmetric theory the coupling is a free parameter. 
Nevertheless this example provides of word of caution with regard to our assertion.

We found two classes of bound states, cyclic and non-cyclic. The cyclic bound states have
color flux that is wrapped completely around the compact transverse space. We were able to
isolate two sequences of such states. Each sequence corresponds to a given state with a
different number of wrappings. As a function of the winding number W the masses have the form
$m^2=A +B/W +C/W^2$.  In the non-cyclic sector we find stringy states as we have in previous
SDLCQ calculations. We find good convergence for the bound states we present as a function of
K.

Finally we would like to note that the symmetries of this approach and those of Cohn, Kaplan,
Katz and Unsal (CKKU)\cite{ckku03} appear to be similar. The formulations are totally different,
and these authors consider a two--dimensional discrete spacial lattice as well as extended
supersymmetry. Nevertheless there are some similarities. As we have noted several times we have
color conservation at each lattice site, thus the symmetry group is $U(N_c)^{N_{sites}}$ similar
to CKKU. We have enforced translation invariance for
this discrete lattice with $N_{sites}$ sites; therefore, there is a
$Z_{N_{sites}}$ symmetry similar to one found by CKKU for their two dimensional lattice.
Finally, in this theory there is an orientation symmetry for the trace which is a $Z_2$ symmetry
also similar to CKKU. In addition CKKU have some $U(1)$ symmetries which we seem to be missing.
This may be related to the fundamentally different way chiral symmetry is treated on the
light cone\cite{bpp98}. Another similarity appears to be the relation between the number of
supersymmetries and the number of fermions on a site. Both approaches have one fermion on a site
and one supersymmetry. 

Numerically this calculation was done using our Mathematica code on a Linux workstation. This
was very convenient for our first attempt at a supersymmetric formulation of a transverse
lattice problem. The trade off is that it limits significantly how far we can go in resolution
and in the number of sites.  Our current
$C++$ code can be modified to handle this problem and will allow a significantly increased
resolution. We should also be able to handle the problem of two transverse dimensions with this
code.  These appear to be fruitful directions for future research.

\section*{Acknowledgments}
This work was supported in part by the U.S. Department of Energy. The authors would like to
acknowledge very useful conversations with Brett Van de Sande as well as the help and
assistance of John Hiller.

\end{document}